# CdTe and CdZnTe Crystal Growth and Production of Gamma Radiation Detectors

Uri Lachish, *guma* science, P.O. Box 2104, Rehovot 76120, Israel

urila@internet-zahav.net

**Abstract**

Bridgman CdTe and CdZnTe crystal growth, with cadmium vapor pressure control, is applied to production of semiconductor gamma radiation detectors. Crystals are highly donor doped and highly electrically conducting. Annealing in tellurium vapors transforms them into a highly compensated state of high electrical resistance and high sensitivity to gamma radiation. N-type detectors, equipped with ohmic contacts, and a grounded guard ring around the positive contact, are not sensitive to hole trapping. Conductivity control, by the doping level, optimizes the detector operation by trade-off between electrons' lifetime and electrical resistance. Gamma spectra of single detectors and detector arrays are presented. Detector optimization and gamma detection mechanisms are discussed.

## 1. Introduction

CdTe and CdZnTe (CZT) semiconductor gamma detectors have been made traditionally from crystals that grow by the traveling heater method (THM) [1 - 2], or by the method of high pressure Bridgman (HPB) [3 - 6].

In THM, chlorine doped crystals grow from tellurium solution. The ratio of cadmium to tellurium within the solution determines its melting point, therefore, also the temperature of crystal growth, that is usually adjusted in the range $650^0$ - $700^0$ C. High quality crystals and detectors have been made in this way. However, the low growth temperature limits the growth rate to typically few millimeters a day. The slow growth rate is the main disadvantage of this method.

In high pressure Bridgman (HPB), crystals grow from a melt of nearly equal quantities of cadmium and tellurium, with small cadmium excess. The cadmium excess generates high vapor pressure that requires growth furnace of special design. The crystal grows at high temperature, above $1100^0$ C, at a high growth rate of few millimeters per hour. Zinc addition increases the crystal's band gap and electrical resistivity, therefore, it reduces the detector's spectral broadening by dark current noise. HPB yields high quality detectors and detector arrays, however, the crystal uniformity is limited, and the detectors' yield is low. Control of the high pressure system is inconvenient.

CdTe and CdZnTe grow by standard low pressure Bridgman techniques that have been developed and perfected to production of substrates for infra-red detectors.

However, these techniques were not considered suitable to production of detector grade material.

This work describes the original application of low pressure Bridgman crystal growth techniques to produce high quality CdTe and CdZnTe gamma detectors and detector arrays.

Extensive work has been done on producing monolithic gamma detector arrays equipped with common negative contact and positive contact array. Spectra, of single pixels in these arrays, achieve optimal spectral resolution by adjusting the gamma charge collection time (shape time) to the electron transition time from contact to contact. Thus, the detector circuit collects only the electrons' contribution to the signal, and the spectral response is not deteriorated by the missing charge of the holes. Similar data have been observed in single detectors, equipped with positive contacts surrounded by grounded [7] or voltage biased [8] guard-rings.

Insensitivity to hole trapping enables flexible detector design and adjustment of its physical parameters to optimal operation.

## 2. Crystal growth

CdTe crystals grow by the low pressure Bridgman technique in a closed evacuated ampoule with small tellurium excess [9]. The three material phases, solid, liquid and gas, coexist during the growth, and the pressure in the ampoule is nearly equal to the tellurium vapor pressure at the growth temperature. Under these conditions the crystal contains excess tellurium atoms that during cool down come out as precipitates, mainly at dislocations and grain boundaries.

The crystal quality improves and precipitates are avoided by applying the modified Bridgman technique [9 - 12]. In this procedure there is small cadmium excess in the ampoule. During crystal growth one ampoule end is kept at lower temperature that determines a nearly atmospheric constant vapor pressure in the system. The growth process involves continuous material transfer between the three phases. The constant vapor pressure keeps constant liquid composition, and balanced amounts of cadmium and tellurium within the crystal.

Figure 1 presents the horizontal and vertical versions of this technique. In both cases the crystal grows from the melt by moving it along a region with temperature gradient that extends from above to below the melting point. The growth may proceed by mechanically moving the ampoule, or, by moving the heating furnace. In recent systems the furnace consists of many heating zones and a computer controls the temperature profile. The computer shifts the profile electronically and there are no mechanically moving parts within the furnace [10- 12].

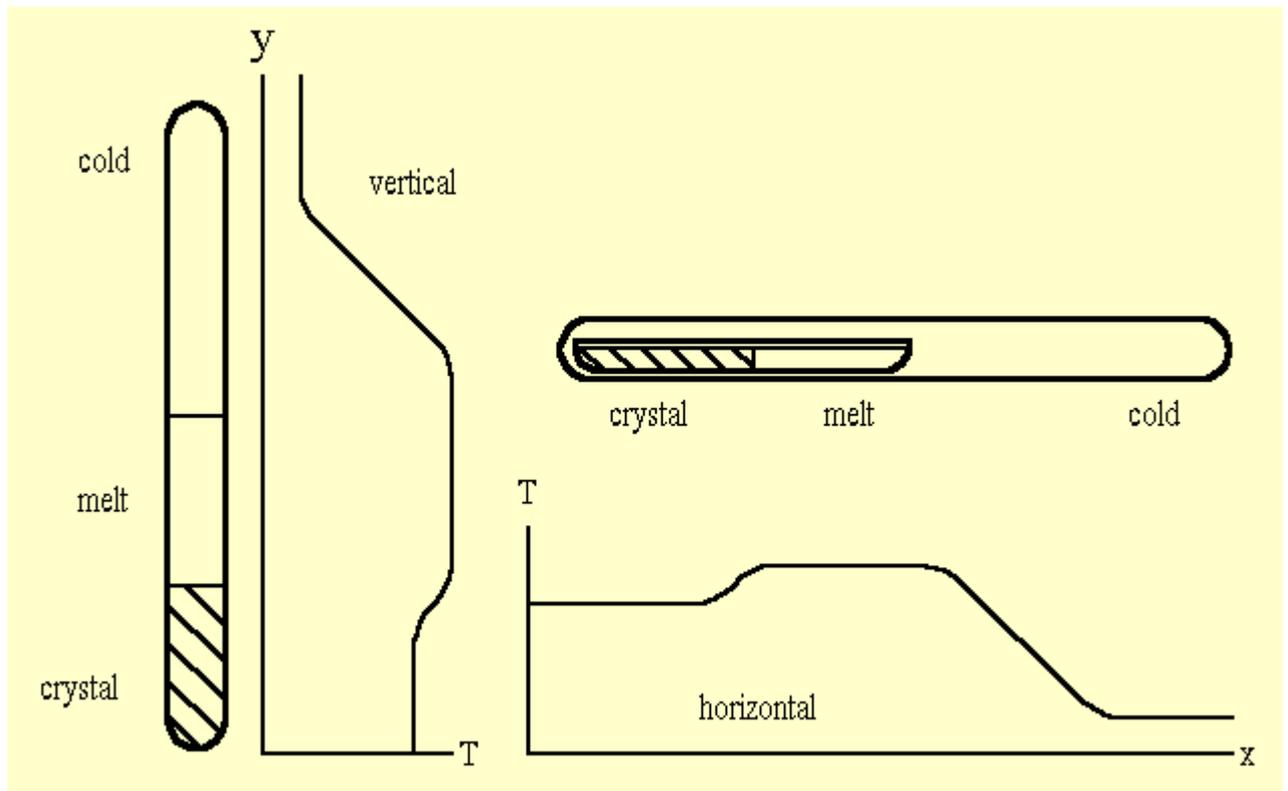

**Figure 1.** CdTe Crystal growth by vertical and horizontal Bridgman with cadmium vapor pressure control. Constant vapor pressure keeps constant liquid composition, and balanced amounts of cadmium and tellurium within the crystal.

The growth procedure starts by melting the separate cadmium and tellurium loads, under hydrogen atmosphere, in order to remove oxygen from the system [11 - 12]. The materials are then brought into contact and heated until they react and produce CdTe, or, CdZnTe if zinc is also present. The material is further heated above the melting point, and crystal growth starts after homogenizing period. Good quality crystals, practically free of precipitates, grow when the cold end temperature is about $300^0$ C below the melting point.

After crystal growth is complete it is cooled down slowly, in order to avoid internal stress and damage by temperature gradients within it. Control of the ampoule's cold end temperature is important during the cool down period, in order to have good crystal quality. The crystal quality is maintained when the difference temperature, between the crystal and cold end, stays constant during this stage. The excess cadmium in the system condenses separately and crystallizes at the coldest point in the ampoule.

**3. Control of the electrical conductivity**

Annealing CdTe in cadmium vapors introduces excess cadmium atoms into the crystal. Annealing in tellurium vapors introduces tellurium atoms to the crystal, or, removes cadmium atoms from it. Either way, the result will be excess tellurium atoms within the crystal. Cadmium excess in pure CdTe transforms it into a medium resistivity, $10^5$ - $10^6$ ohm cm, n-type material. Tellurium excess transforms it to $10^3$ - $10^5$ ohm cm p-type material.

High resistance CdTe may be produced by proper annealing. However, pure CdTe is not gamma active. Addition of a donor dopant activates the material.

Crystals with uncompensated donors are highly conductive. Annealing with tellurium vapors transforms the donors into a highly compensated state. The crystal transforms into a state of high electrical resistance and high sensitivity to gamma radiation. The free donor concentration in the compensated state, that determines the crystal's resistance, is many orders of magnitude lower than the overall dopant concentration. However, the two quantities are related, so that increasing the overall dopant level will increase also the free donor level. Therefore, the crystal conductivity is controllable by adjustment of the overall dopant level in it.

Trivalent atoms, indium or aluminum, act as donors by replacing cadmium atoms in the crystal lattice, and halide atoms, chlorine, by replacing tellurium. However, when CdTe grows with too much chlorine, then separate $CdCl_2$ crystal will grow in the ampoule. Limited chlorine amount can enter into the crystal in the high temperature Bridgman growth, therefore, it is not a suitable dopant for conductivity control. It is possible that chlorine doping is adequate to THM crystals because higher amounts can enter into the crystal at the THM lower growth temperature.

Indium doping yields even dopant distribution within the crystal from end to end. Tellurium annealing at low enough temperature does not damage the crystal and precipitates do not appear. Diffusion processes are slow at low temperatures. Therefore, cutting the crystal into wafers, prior to annealing, will reduce the annealing time [13].

A series of iterations of crystal growth experiments with different doping levels, followed by subsequent annealing, yields optimal crystal composition of the required electrical conductivity.

In summary, doped crystals come out highly conducting under optimal crystal growth conditions. Annealing in tellurium vapors transforms them into a compensated state of high electrical resistivity and high sensitivity to gamma radiation. The electrical conductivity, in this state, is controllable by adjustment of the doping level.

High resistance material can be made in a single process of crystal growth, by proper reduction of the cadmium vapor pressure during the cool down stage. However, better detectors are achieved by dividing the process into separate steps of growth and annealing, adjusted to optimal crystal quality and electronic performance.

**4. Device design**

Figure 2 shows schematic presentation of trapping levels, within the forbidden gap, between the valence and conduction bands of a semiconductor. Levels that are located above the Fermi level act as electron traps, and levels below it act as hole traps. Shift of the Fermi level, up or down, will decrease the amount of trap levels of one type on expense of the other type. Therefore, it will also increase the lifetime of one charge carrier on expense of the other carrier.

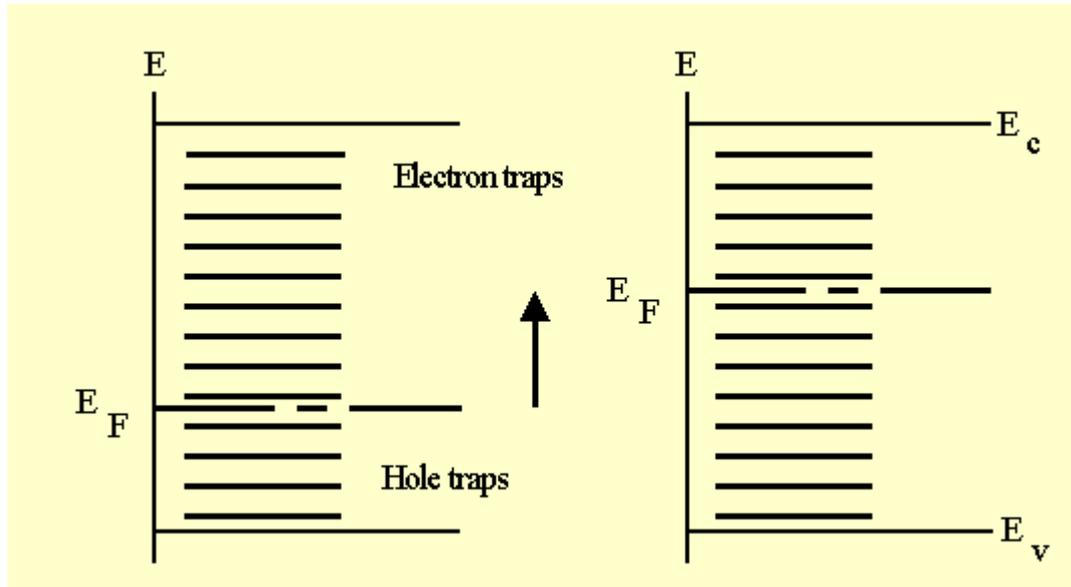

**Figure 2.** Schematic presentation of trap levels within the forbidden gap of a semiconductor. Fermi level push-up inactivates electron traps, that shift below it, and increases the electrons' lifetime on expense of the holes. The level shift also increases the electrical conductivity towards n-type. Trade-off between the electrons' lifetime and electrical resistance optimizes the Fermi level position and detector operation.

Push up of the Fermi level inactivates electron traps that go below it. Therefore, it increases the electrons' lifetime on expense of the holes, and improves the performance of detectors that are not sensitive to hole trapping. However, this shift towards n-type conduction also increases the conductivity, and the higher leakage current will increase the noise and broaden spectral lines.

The detector operation is optimized by adjustment of the electrical conductivity, by trading off electrons lifetime with electrical resistance. In this way, the Fermi level is optimally positioned within the forbidden gap. In practice, the detector operates best by adjusting the n-type conduction to somewhat above its lowest possible value.

Figure 3 is a schematic layout of trapping levels presenting the compensation process in more detail.

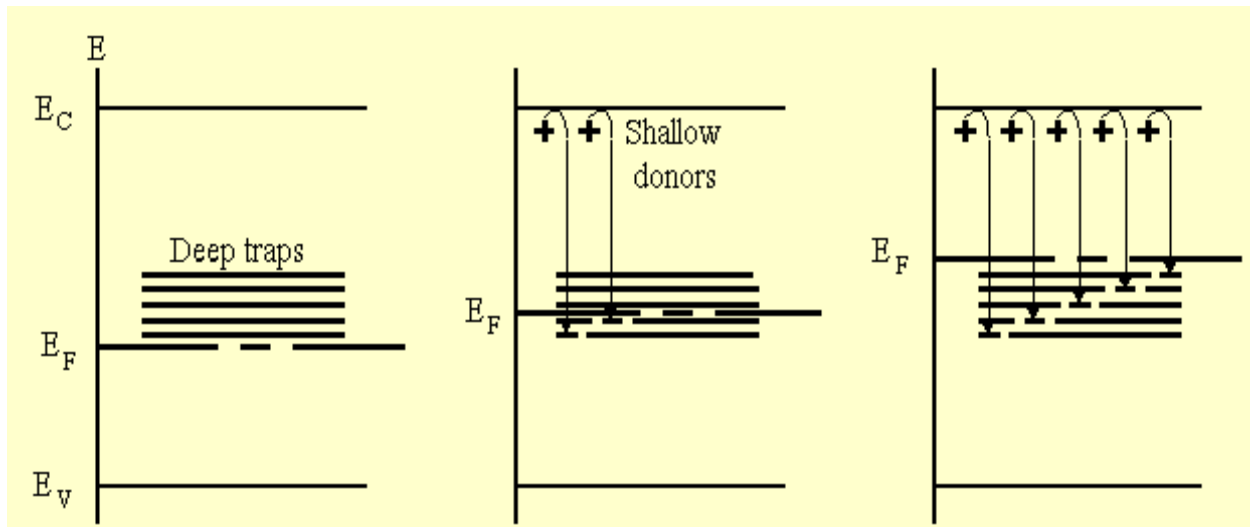

**Figure 3.** Inactivation of deep electron traps by addition of donor dopants. Trivalent donor, such as indium, replaces bivalent cadmium atom within the crystal lattice, and the extra electron falls into the deep trap, leaving behind it a shallow ionized donor.

A narrow band of deep electron trapping levels, located somewhere within the middle of the forbidden gap, trap the gamma generated electrons and shorten their lifetime. When a donor dopant, such a trivalet indium, replaces a bivalent cadmium atom within the crystal lattice, then the extra electron will fall into a deep trap, leaving behind it an ionized shallow donor. Addition of more donors shifts up the Fermi level from below the trapping band to somewhere within it. The optimal donor concentration is achieved when nearly all the deep traps become occupied and replaced by ionized shallow donors, and the Fermi level shifts to just above the deep trapping band. There is no need of short distance interaction between the dopant donor atom and the deep trap.

In the undoped crystal the Fermi level is pinned by the deep trap band, and addition of fairly large amount of donors leads to only small level shift. The optimal operating point is achieved when the band becomes nearly occupied and the donor concentration unpins the Fermi level position.

Gold or platinum contacts are applied to CdTe, or CdZnTe, by evaporation, sputtering, or chemical deposition. Indium evaporation achieves good ohmic contacts to n-type material. Indium contacts require thermal annealing that is limited by its melting point of $156^0$ C. Aluminum cap layer allows higher annealing temperatures.

## 5. Results

Figure 4 shows a standard measurement scheme for gamma detector characterization. A gamma source irradiates a monolithic detector array through the negatively biased common contact. The tested pixel is connected to the input of a preamplifier (Ortec 142A) while its high voltage input is grounded. The preamplifier is connected to a spectroscopy amplifier (Ortec 572), then to a multi channel analyzer (MCA). The response time of the spectroscopy amplifier is adjusted to $0.5 \times 10^{-6}$ s.

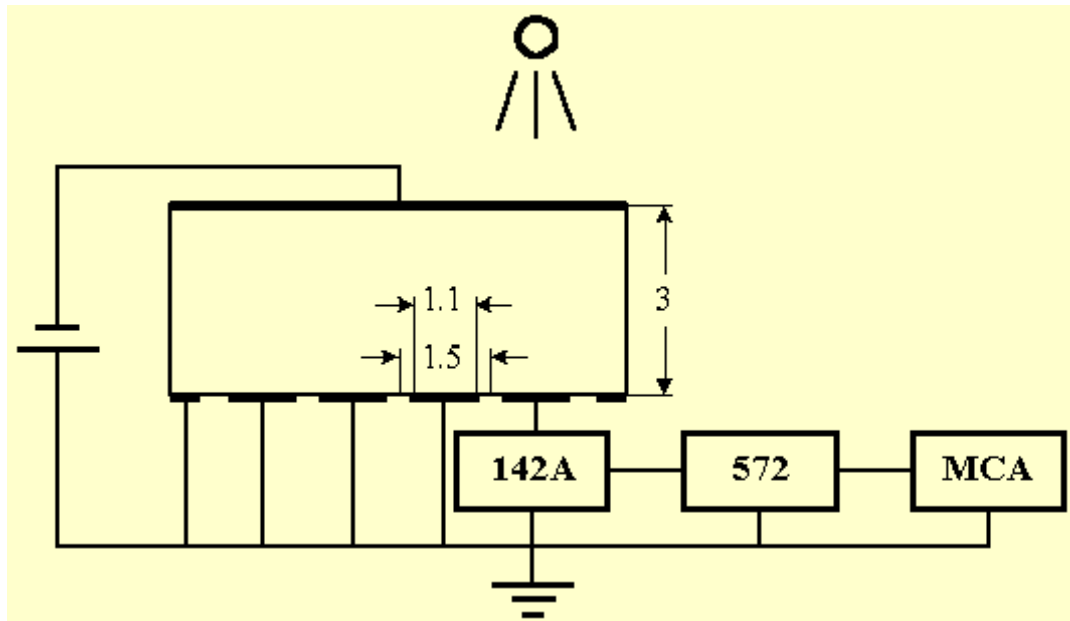

**Figure 4.** Measurement scheme of a semiconductor detector array. Each pixel is tested separately, while all the other pixels and the guard ring are grounded. The device dimensions are given in mm.

Figure 4 also shows the dimensions in millimeters of a CdTe 4 x 4 pixels monolithic detector array, equipped with indium contacts. The electrical resistivity is $5 \times 10^8$ ohm cm.

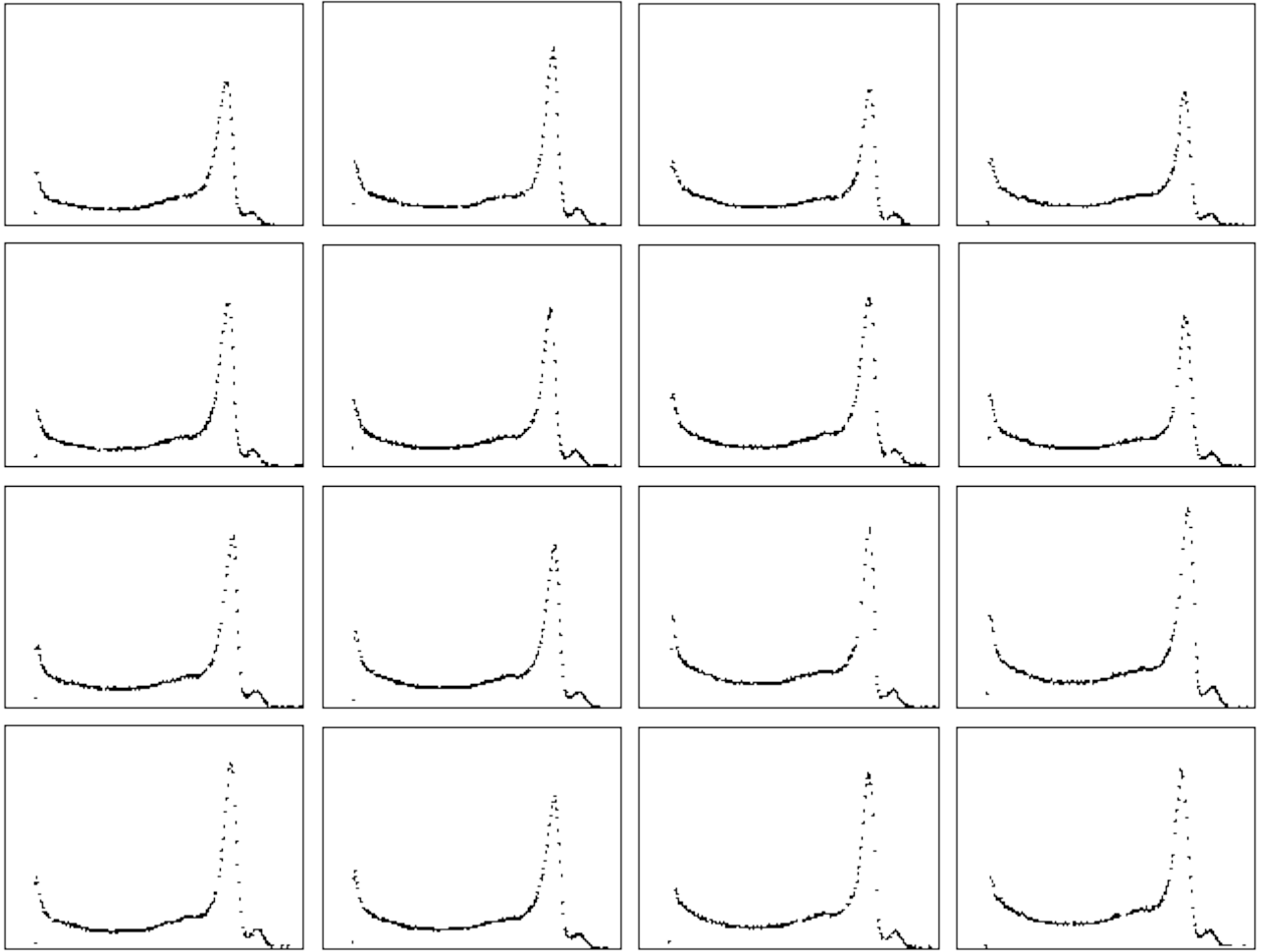

**Figure 5.** $^{57}$Co spectra of a CdTe 4 x 4 detector array, for bias voltage -200 v and shape time 0.5 x 10$^{-6}$ s.

Figure 5 shows the 16 pixels' spectra measured with a $^{57}$Co source (122 kev) for a bias voltage of -200 volts. Each pixel is tested separately while the source is located above it, and all the other pixels and the guard ring are grounded. All the pixels are active and the figure indicates uniform spectral response.

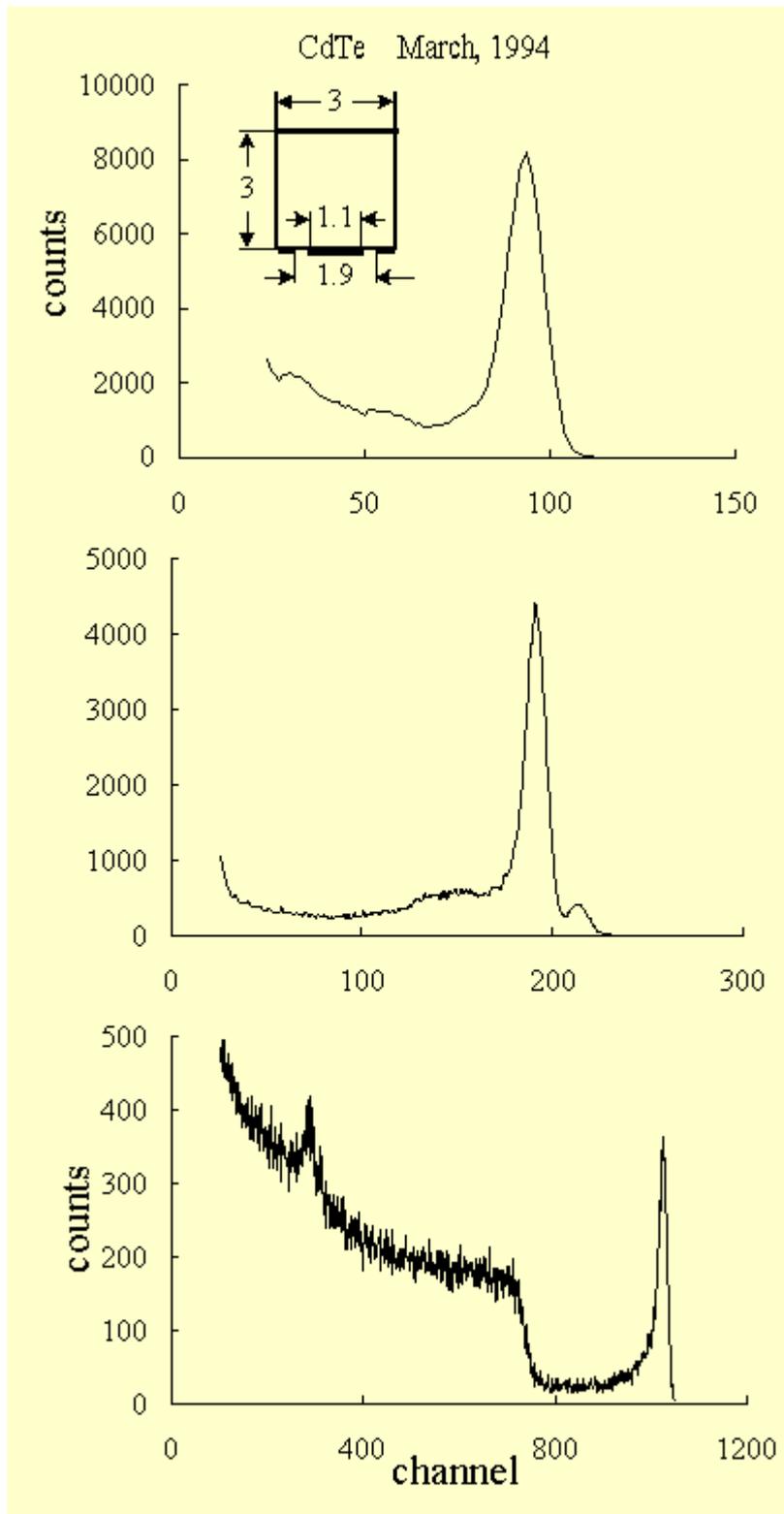

**Figure 6.** $^{241}$Am, $^{57}$Co and $^{137}$Cs spectra of a CdTe detector, for bias voltage -200 v and shape time 0.5 x 10$^{-6}$ s. The detector dimensions in mm are given in the insert. FWHM are 13.5%, 7.0% and 2.2% respectively.

Figure 6 shows the $^{241}$Am (59.5 kev), $^{57}$Co (122 kev), and $^{137}$Cs (662 kev) spectra of a single detector with a grounded guard-ring around the positive contact. The detector dimensions are in the figure. The bias voltage and shape time, are as in figure 5.

The spectral FWHM are 13.5%, 7.0% and 2.2% respectively. The relatively wide and symmetric $^{241}$Am line indicates current noise broadening that is due to the relatively low detector electrical resistance.

The electron transition time from contact to contact, for detector thickness $d = 3$ mm, bias voltage $V = -200$ volts, and mobility $\mu = 10^3$ cm$^2$ v$^{-1}$ s$^{-1}$, is:

$$\tau = d^2/\mu V = 0.45 \times 10^{-6} \text{ s} \quad (1)$$

The time $\tau$ is similar to the instrumental charge collection (shape) time. Therefore, the detector circuit mainly collects the electrons' contribution to the signal, since during this period the holes hardly move at all. The good line resolution of $^{137}$Cs, observed in figure 6 for the nearly uniform 662 kev excitation, indicates that the missing contribution of holes does not deteriorate the signal.

A narrow grounded guard-ring must surround the positive contact, in order to have good spectral resolution. In segmented contact a guard ring should surround the positive array, and all pixels should be connected to preamplifiers

Table-1 summarizes data on the effect of dopant level observed with many 3 mm thick CdTe detectors.

| dopant level | $\rho$ ($\Omega$ cm) | $10^{-3} \mu \tau$ (e) (cm$^2$/volt) | performance |
|---|---|---|---|
| undoped | $> 10^9$ |  | not active |
| under-doped | $> 10^9$ | $< 0.5$ | 3 x bias |
| properly doped | $5 \times 10^8$ | 1 - 2 | good |
| over-doped | $< 10^7$ | $>10$ | poor |

**Table-1.** Effect of the dopant level on the detector performance.

Undoped material is not gamma active and only n-type doped material is active. Under-doped detectors operate reasonably at high bias voltage, that is about three times as much as optimized detectors. Properly doped detectors have good spectral resolution, as observed in figures 4, 5. However, the resistivity of the n-type material is lower than that of p-type detectors made by other methods, and the spectral resolution is current limited. Over-doped detectors have very high $\mu \tau$ (e), but the performance is very poor since the electrical resistance is very low.

## 6. Models of detection mechanism

There are two models that account for the observed insensitivity of the pixels' spectra to the hole contribution to the signal.

According to the small pixel effect the current in a single pixel circuit, of a detector array, depends on the distance from the pixel. Electrons flow from the point of gamma excitation, and contribute to the charge signal of a single pixel circuit when they arrive near to it. Holes flow towards the common negative contact, and their current contribution is distributed over a number of pixels. Therefore, excluding the hole contribution from a single pixel signal, will have a marginal effect of adding a low energy tail to the spectral line.

The small pixel theory (Barrett et al. [14] figure 1, Eskin et al. [15] figure 11) predicts time dependent signal during the flow of electrons from the negative to the positive contact. The signal in the detector circuit starts to build up fast only when the electrons arrive near to the positive contact. Therefore, the spectral resolution improves by reducing the instrumental shape time, below the electron transition time from contact to contact, to fit the fast signal rise-up. The shorter shape time filters out the slow contribution of charge flow near the negative contact. Thus, it removes the low energy spectral tail.

The model of ohmic contact effect considers single detectors, or detector arrays, equipped with ohmic contacts. It suggests that Poisson and the continuity equations require that the flow of gamma generated charge induces additional electrons flow, from the negative contact towards the holes [16 - 17]. Contacts to n-type material, observed ohmic by the leakage current, are also ohmic for the gamma induced recombination current.

Both models predict fast charge signal in the pixel's circuit, and insensitivity to the holes' movement. They differ in the time dependent current that follows gamma excitation. According to the ohmic contact model, afterglow current will follow the charge recombination current.

## 7. Summary and Conclusions

CdTe and CdZnTe crystals grow by the Bridgman method, with cadmium vapor pressure control. The highly donor doped crystals are highly electrically conducting. Annealing in tellurium vapors transforms crystal wafers into a highly compensated state of high electrical resistance and high gamma sensitivity, controlled by the doping level.

Detectors equipped with ohmic contacts, and a grounded guard-ring around the positive contact, have fast charge collection time. They are not sensitive to hole trapping.

The electrical conductivity determines the Fermi level position within the semiconductor band gap. Conductivity adjustment, in detectors insensitive to hole trapping, optimizes the detector operation by trade-off between electrons' lifetime and electrical resistance. The detector performance improves by increasing the conductivity above its minimal value.

Detectors insensitive to hole trapping are uniform and applicable to production of large area detector arrays.

So far the production of medium and large scale detector arrays for commercial applications has been hindered by low yield of material quality for detectors. However, by a proper technology it is possible to grow cystals with good yield of detector grade material for gamma detection.

On the net: 22nd March, 1998. Revised: February 2000. Section 4 revised: November 2000. Conclusion refined: November 2006.

by the author:

1. "Driving Spectral Resolution to the Noise Limit in Semiconductor Gamma Detector Arrays", IEEE Trans. Nucl. Sci., Vol 48(3), pp 520 - 523, June (2001).